\documentclass[aps,twocolumn,superscriptaddress,showpacs]{revtex4}

\usepackage{amssymb}
\usepackage{amsmath}
\usepackage{graphicx}
\usepackage[normalem]{ulem}
\usepackage[dvips]{color}

\usepackage[normalem]{ulem} 
\usepackage[dvips]{color} 
\renewcommand\sout{\bgroup \color{red} \ULdepth=-.5ex \ULset}

\def\esym{$E_{sym}(\rho)$~}
\def\rpi {$\pi^-/\pi^+$~}
\def\es0{$E_{sym}(\rho_0)$~}
\begin{document}

\title{Influence of neutron-skin thickness on $\pi^{-}/\pi^{+}$ ratio in Pb+Pb collisions}
\author{Gao-Feng Wei}\email[Email address: ]{wei.gaofeng@foxmail.com}
\affiliation{Department of Physics and Astronomy, Texas A$\&$M
University-Commerce, Commerce, TX 75429-3011, USA}
\affiliation{School of Physics and Mechatronics Engineering, Xi'an University of Arts and Science, Xi'an, 710065, China}
\author{Bao-An Li}\email[Corresponding author's email address: ]{Bao-An.Li@tamuc.edu}
\affiliation{Department of Physics and Astronomy, Texas A$\&$M
University-Commerce, Commerce, TX 75429-3011, USA}
\affiliation{Department of Applied Physics, Xi'an Jiao Tong University, Xi'an 710049, China}
\author{Jun Xu}\email[Email address: ]{xujun@sinap.ac.cn}
\affiliation{Shanghai Institute of Applied Physics, Chinese Academy of Sciences, Shanghai 201800, China}
\author{Lie-Wen Chen}\email[Email address: ]{lwchen@sjtu.edu.cn}
\affiliation{Department of Physics and Astronomy and Shanghai Key Laboratory for Particle Physics and Cosmology,
Shanghai Jiao Tong University, Shanghai 200240, China}
\affiliation{Center of Theoretical Nuclear Physics, National Laboratory of Heavy Ion Accelerator, Lanzhou 730000, China}

\begin{abstract}
Within an isospin- and momentum-dependent transport model (IBUU11)
using as an input nucleon density profiles from Hartree-Fock
calculations based on a modified Skyrme-like (MSL) model, we study
the influence of the uncertainty of the neutron skin thickness on
the $\pi^{-}/\pi^{+}$ ratio in both central and peripheral Pb+Pb
collisions at beam energies of 400 MeV/nucleon and 1000 MeV/nucleon.
Within the current experimental uncertainty range of neutron skin in
$^{208}$Pb,  while the neutron skin effect on the \rpi ratio is
negligible in central reactions at both energies, it increases
gradually with increasing impact parameter and becomes comparable
with or even larger than the symmetry energy effect in peripheral
collisions especially at 400 MeV/nucleon. Moreover, we find that
while the \rpi ratio is larger with a softer \esym in central
collisions, above certain impact parameters depending on the size of
the neutron skin, a stiffer \esym can lead to a larger \rpi ratio as
most of the pions are produced at densities below the saturation
density in these peripheral reactions. Therefore, a clear impact parameter
selection is important to extract reliable information about the
\esym at suprasaturation densities (size of neutron skin)
from the $\pi^-/\pi^+$ ratio in central (peripheral) heavy-ion collisions.
\end{abstract}


\maketitle

\section{Introduction}\label{introduction}
The density dependence of nuclear symmetry energy \esym affects not
only the structure of nuclei and neutron stars, such as the neutron
skin in heavy nuclei and radii of neutron stars \cite{Ste05a,Jim13}, but also their
reaction dynamics, such as particle production in heavy-ion
collisions \cite{LiBA98,LiBA01b,Dan02a,Bar05,LCK,Lynch09,Trau12} and emission of gravitational waves in spiraling neutron star binaries \cite{Far12}. Thus, combining information from various laboratory
experiments and astrophysical observations has the promise of
mapping out accurately the currently still poorly known \esym in a broad
density range \cite{Ste05b,LiBA06a}.
In particular, effects of various observables from the size of neutron skin, pygmy dipole resonance, and dipole polarizability in heavy nuclei (see, e.g., Refs. \cite{Cen09,Rei10,Roc11,RCNP}) to the neutron/proton \cite{LiBA97a} and \rpi \cite{Bao02}
ratios in heavy-ion collisions have been found to be sensitive to the \esym at different densities.
While significant progress has been made in constraining the \esym mostly around
saturation density $\rho_0$ \cite{Tsang12}, much more work needs to be done to
better determine the \esym at both subsaturation and suprasaturation
densities. For an overview of the latest status of the field, we
refer the reader to the web pages of the 2013 International
Collaboration in Nuclear Theory Program on Nuclear Symmetry Energy
\cite{ICNT} and The Third International Symposium on Nuclear
Symmetry Energy (Nusym13) \cite{Nusym13} as well as the very recent EPJA Topical Issue on Nuclear Symmetry Energy~\cite{EPJA}.

The size of neutron skin has long been identified as one of the most
promising observables to fix the \esym around 0.1 fm$^{-3}$, see,
e.g., Refs. \cite{Bro00,Typ01,Hor01,Fur02,Lwchen1,Lwchen2}. However,
the available data obtained mostly from hadronic probes suffer from
large uncertainties. For example, the experimentally measured size
of neutron skin in $^{208}$Pb currently ranges from about $0.11\pm 0.06$ fm
from $\pi^+$-Pb scattering~\cite{frie} to $0.33^{+0.16}_{-0.18}$ fm from the PREX-1 experiments
using parity violating e-Pb scattering \cite{Prex1}.
For a recent review, see, e.g., Ref. \cite{Pawel13}. In
particular, it was shown very recently within a relativistic
mean-field model \cite{Farrooh} that a neutron skin for $^{208}$Pb
as thick as $0.33+0.16$ fm reported by the PREX-I experiment \cite{Prex1}
can not be ruled out although most other studies have reported much
smaller average values albeit largely overlapping with the PREX-I result within error bars.
This situation has stimulated the renewed interest
to experimentally measure model-independently sizes of neutron skins
in medium and heavy nuclei. In this regard, it is interesting to note that the approved CREX and PREX-II experiments at JLab \cite{CREX} are expected to provide more accurate values for the neutron skin thickness in both $^{48}$Ca and $^{208}$Pb.

On the other hand,  the \rpi ratio in heavy-ion collisions near the pion production
threshold \cite{Bao02} is among the most promising tracers of nuclear symmetry energy at suprasaturation densities based on transport model simulations by several groups.
Unfortunately, the theoretical results from different models are still inconsistent with each other, and
conclusions from comparisons with very limited data available remain
rather controversial \cite{Xiao09,Feng02,Xie13,Xu13,HJ13}.
This situation demands certainly collective
efforts by the community, among other things, to quantify
theoretical uncertainties associated with various model assumptions
and input parameters in transport model studies of heavy-ion
collisions. As a useful step towards this goal, we investigate
effects of the uncertainties of the neutron skin thickness on
extracting information about the high-density symmetry energy using
the $\pi^{-}/\pi^{+}$ ratio in heavy-ion collisions within the latest version of
an isospin and momentum dependent transport model (IBUU11)
\cite{Ouli}. Because of the diverse predictions using various interactions and
many-body theories, the neutron skins for colliding nuclei used in
transport models vary significantly. We investigate here how
the size of neutron skin may affect the \rpi signal of the \esym
from central to peripheral heavy-ion collisions. This is an important issue because the charged pion ratio has the dual sensitivity to both the size of neutron-skin and the density dependence of nuclear symmetry energy.
The two are obviously interconnected. In fact, soon after
the high-energy radioactive beams become available some 25 years
ago, the pion production in peripheral nuclear reactions was
proposed as a sensitive measure of the size of neutron skin in rare
isotopes \cite{Tel87,Lom88,Bao91}. It is thus necessary to examine
the influence of the uncertainty of neutron skin thickness on the
information of \esym we may extract from studying the \rpi ratio in
heavy-ion collisions. As an example, we study the \rpi ratio in
Pb+Pb collisions from central to peripheral impact parameters at
beam energies of 400 MeV/nucleon and 1000 MeV/nucleon. We will see
that while the neutron skin effect on the \rpi ratio is found
negligible in central reactions at both energies, it increases
gradually with increasing impact parameter and becomes comparable
with or even larger than the symmetry energy effect in peripheral
collisions especially at 400 MeV/nucleon. Moreover, although the
\rpi ratio is larger with a softer \esym in central collisions, we
found that above certain impact parameters depending on the size of
the neutron skin, a stiffer \esym can lead to a larger \rpi ratio as
most of the pions are created at densities below the saturation
density in these peripheral reactions.

\section{nuclear symmetry energy and initialization in IBUU11}
\begin{figure}[h]
\centerline{\includegraphics[scale=0.35]{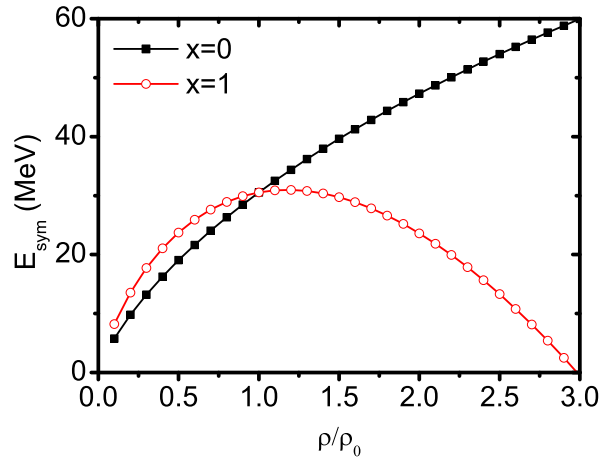}} \caption{(Color
online) The density dependence of the symmetry energy.} \label{Esym}
\end{figure}

This study is carried out using the isospin dependent
Boltzmann-Uehling-Uhlenbeck (IBUU) transport model \cite{IBUU} of
version IBUU11 \cite{Ouli}. To ease the following discussions, we
briefly describe here the isospin and momentum dependent interaction (MDI)
used in the IBUU11 and the nucleon density
profiles with different sizes of neutron skins used in initializing
the two colliding nuclei. The momentum dependence of both the
isoscalar \cite{GBD87,PDG88,MDYI90,Dan00,Greco99} and isovector
\cite{IBUU,Das03,Chen04,Rizzo04} parts of the nuclear interaction
is important in understanding not only many phenomena in
intermediate-energy heavy-ion collisions but also thermodynamical
properties of isospin-asymmetric nuclear matter \cite{Xuj07a,Xuj07b,Xuj08}.
The MDI mean-field potential for a nucleon with momentum $\vec{p}$ and
isospin $\tau$ can be written as \cite{Das03}
\begin{eqnarray}
U(\rho,\delta ,\vec{p},\tau ) &=&A_{u}(x)\frac{\rho _{-\tau }}{\rho _{0}}%
+A_{l}(x)\frac{\rho _{\tau }}{\rho _{0}}  \notag \\
&+&B(\frac{\rho }{\rho _{0}})^{\sigma }(1-x\delta ^{2})-8\tau x\frac{B}{%
\sigma +1}\frac{\rho ^{\sigma -1}}{\rho _{0}^{\sigma }}\delta \rho
_{-\tau }
\notag \\
&+&\frac{2C_{\tau ,\tau }}{\rho _{0}}\int d^{3}p^{\prime }\frac{f_{\tau }(%
\vec{p}^{\prime })}{1+(\vec{p}-\vec{p}^{\prime })^{2}/\Lambda ^{2}}
\notag \\
&+&\frac{2C_{\tau ,-\tau }}{\rho _{0}}\int d^{3}p^{\prime }\frac{f_{-\tau }(%
\vec{p}^{\prime })}{1+(\vec{p}-\vec{p}^{\prime })^{2}/\Lambda ^{2}}.
\label{MDIU}
\end{eqnarray}%
In the above, $\rho=\rho_n+\rho_p$ is the nucleon number density and
$\delta=(\rho_n-\rho_p)/\rho$ is the isospin asymmetry of the
nuclear medium; $\rho_{n(p)}$ denotes the neutron (proton) density,
the isospin $\tau$ is $1/2$ for neutrons and $-1/2$ for protons, and
$f(\vec{p})$ is the local phase space distribution function. The
expressions and values of the parameters $A_{u}(x)$, $A_{l}(x)$,
$\sigma$, $B$, $C_{\tau ,\tau }$, $C_{\tau ,-\tau }$, and $\Lambda $
can be found in Refs. \cite{Das03,Che05}, and they lead to the
binding energy of $-16$ MeV, incompressibility $212$ MeV for
symmetric nuclear matter, and symmetry energy $E_{sym}(\rho_0)=30.5$
MeV at saturation density $\rho_0=0.16$ fm$^{-3}$, respectively.

The MDI mean-field potential comes from Hartree-Fock calculations
using a modified Gogny force including a zero-range effective
three-body interaction and a finite-range Yukawa-type two-body
interaction \cite{Das03,Che05,Xuj10}. The variable $x$ is introduced
to mimic different forms of the symmetry energy predicted by various
many-body theories without changing any properties of symmetric
nuclear matter and the value of $E_{sym}(\rho_0)$. Shown in Fig.
\ref{Esym} are the \esym with $x=1$ and $x=0$. The density
dependence of \esym around an arbitrary density $\rho$ can be characterized by
the slope parameter there
\begin{equation}
L_u\equiv 3\rho(dE_{sym}/d\rho)_{\rho}
\end{equation}
where $u=\rho/\rho_0$. The softer (stiffer) \esym with $x=1$ ($x=0$) has a value of $L_1=16.4$ (62.1) MeV and $L_2=-100.3 $ (85.6) MeV, respectively.
 Due to the well-known isospin fractionation during heavy-ion collisions, a
larger value of \esym at density $\rho$ generally leads to a smaller isospin
asymmetry there to lower the energy of the system. Since the density reachable depends sensitively on the impact parameter and the \rpi ratio is sensitive to the isospin asymmetry of the participant region,
the dependence of the \rpi ratio on the \esym is expected to be affected by the impact parameter and the size of neutron skin.

\begin{figure}[h]
\centerline{\includegraphics[scale=0.35]{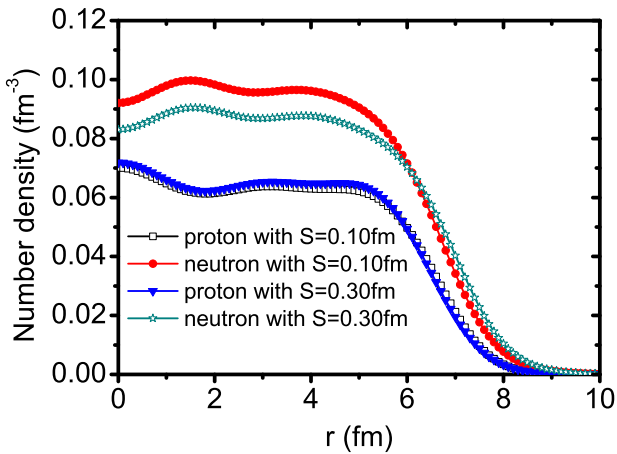}} \caption{(Color
online) The neutron and proton density profiles for $^{208}$Pb with
neutron skin thickness of 0.1 and 0.3 fm, respectively.} \label{Density}
\end{figure}

To initialize transport models, it is necessary to know
the nucleon density profiles for the two colliding nuclei. Ideally,
one would like to use the same nuclear interaction to generate
self-consistently the initial state phase space distributions within
the reaction model itself. Indeed, this was done in some versions of the Quantum Molecular Dynamics models. Another way of initializing nuclei is to use  the Thomas-Fermi (TF) approach using the same interaction as for simulating their subsequent reactions, see, e.g.,
Refs.~\cite{Hol77,Lenk89,Pawel00}.  It is well known that the resulting density profiles
depend on the interaction used. Practically, to our best knowledge, none
of the available reaction models or the TF approach can describe properties of nuclei in their
ground state as good as the microscopic and/or phenomenological
many-body theories for studying nuclear structures. Thus, some reaction simulations use as an input nucleon density profiles predicted by nuclear structure models to initialize nucleons in coordinate space and then the local Thomas-Fermi approximation in momentum space~\cite{IBUU}.
The main problem of this approach is that the initial state may not be the ground state of the interaction used in the subsequent reactions. However, it allows one to easily separate effects on the final observables due to the initial state from those due to the reactions. One of the purposes of this work is to examine the relative effects of the neutron skin in initial nuclei and the symmetry energy at suprasaturation densities reached in central heavy-ion collisions on the charged pion ratio in the final state. The flexibilities of adjusting independently the size of neutron skin of colliding nuclei and the interaction used in the collision simulations are useful for our analyses in this work. Here, we initialize nucleons in phase space using neutron and proton density profiles predicted by
Hartree-Fock calculations based on the MSL model \cite{Che09,Che10}. The size of neutron-skin
$S$ is normally measured by using the difference in the rms radii of neutrons and protons, i.e.,
$S\equiv {<{r_n}^2>}^{1/2}-{<{r_p}^2>}^{1/2}$. Shown in Fig. \ref{Density} are the density profiles corresponding to a neutron skin thickness $S$ of 0.1 and 0.3 fm in $^{208}$Pb. As one expects,
the proton distributions are almost identical, while the neutrons
distribute differently in the two cases considered.

\section{Results and discussions}\label{results}
We now present results of our studies in the following three subsections.
Because the magnitude of isovector potentials is always much smaller
than that of isoscalar potentials during heavy-ion reactions,
isospin effects on reaction observables are generally very small. To
exclude uncertainties of statistical nature from our physical
considerations, we have performed large scale calculations with
$4\times 10^5$ events in each case reported here. Thus, in most
plots the statistical error bars are smaller than the plotting
symbols except in very peripheral reactions.

\begin{figure}[h]
\centerline{\includegraphics[scale=0.35]{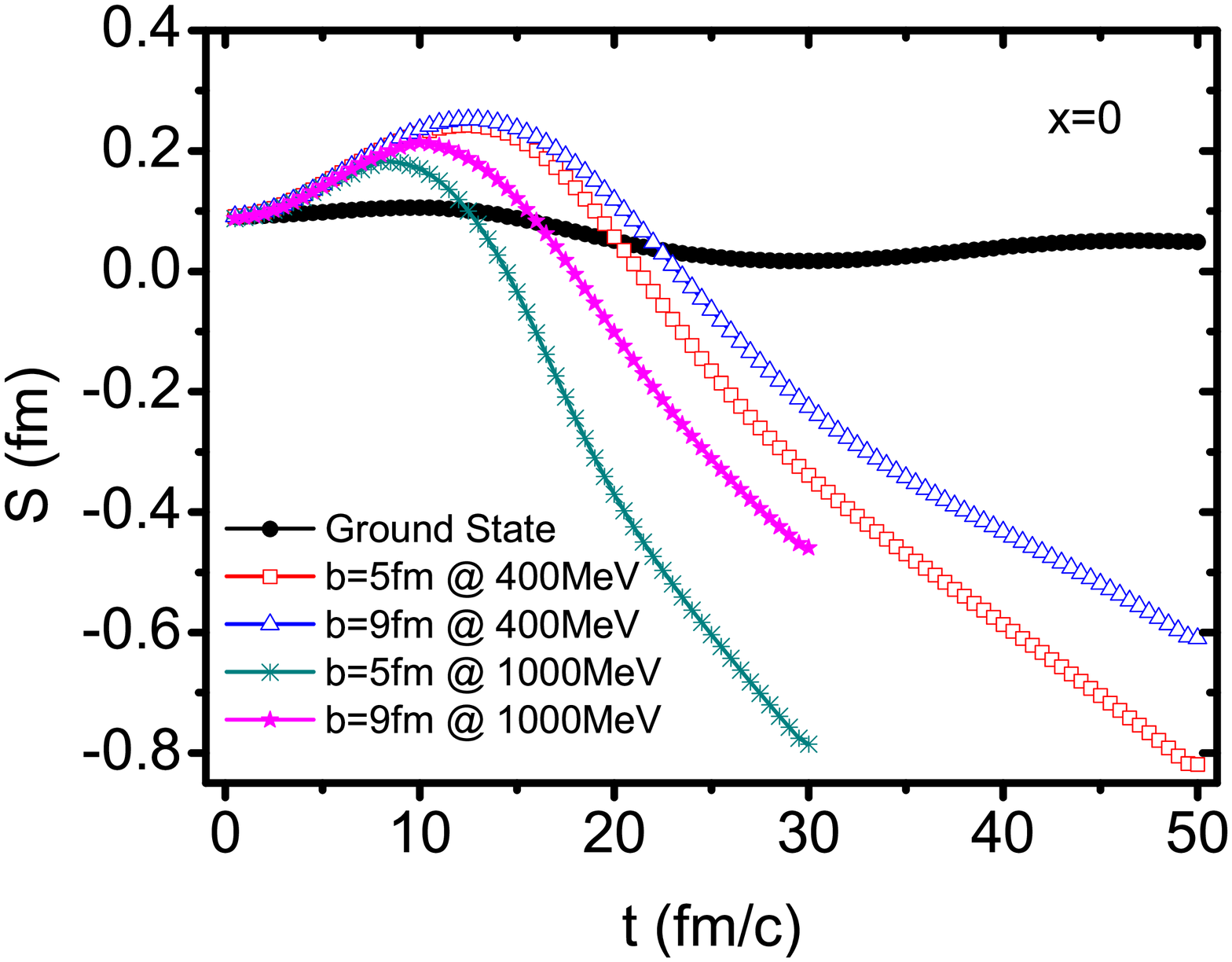}} \caption{(Color
online) Evolution of the size of neutron skin initially set at 0.1 fm in the ground state of $^{208}$Pb, and in the target/projectile
in typical central and peripheral $^{208}$Pb+$^{208}$Pb collisions at beam energies of 400 and 1000 MeV/nucleon, respectively.} \label{tskin}
\end{figure}
\subsection{Evolution of neutron skin and pion production during heavy-ion collisions}
How long does the neutron skin persist at the initial value in the ground state of colliding nuclei and during heavy-ion reactions? How does that compare to the average time to produce a pion? Answers to these questions may help us better understand the interplay between the neutron skin of colliding nuclei and effects of the symmetry energy in nuclear reactions. They may also help evaluate the validity of our model assumptions. To answer these questions, we examine in Fig.\  \ref{tskin} the evolution of the neutron skin initially set at $S$=0.1 fm in the ground state of $^{208}$Pb, and in the target/projectile during typical central (5 fm) and peripheral (9 fm) $^{208}$Pb+$^{208}$Pb collisions at beam energies of 400 and 1000 MeV/nucleon, respectively. It is seen that the initial neutron skin in the ground state lasts longer than the reaction time for the reactions considered, although a long-period and small-amplitude oscillation is visible. This level of stability of the ground state is good enough for the purpose of this study. For simulating nucleus-nucleus collisions, the centers of the two colliding nuclei are initially separated by a distance of $\operatorname{Radius_{target}+Radius_{projectile}+3}$ fm. The two nuclei then approach each other along their Coulomb trajectories before touching. In this case, the initial neutron skin persists for about 5 fm/c. Due to the Coulomb interactions, protons are pushed outward leading to negative $S$ values in the later stage of the reaction.

\begin{figure}[h]
\centerline{\includegraphics[scale=0.37]{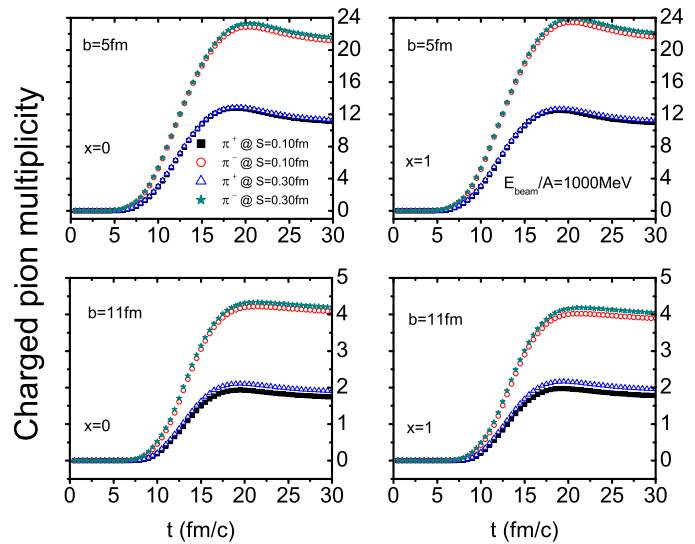}} \caption{(Color
online) Evolution of the charged pion multiplicities in $^{208}$Pb+$^{208}$Pb collisions
at impact parameters of 5 and 11 fm and a beam energy of 1000 MeV/nucleon, respectively.} \label{pmul1}
\end{figure}

Shown in Fig.\ref{pmul1} are the multiplicities of charged pions as a function of time in $^{208}$Pb+$^{208}$Pb collisions
at impact parameters of 5 fm and 11 fm and a beam energy of 1000 MeV/nucleon, respectively. There are clear indications that the multiplicities of both $\pi^-$ and $\pi^+$ increase as the initial size of neutron skin increases from 0.1 fm to 0.3 fm. However, the increase is appreciable only in very peripheral reactions. Unfortunately, effects of the symmetry energy on the pion multiplicities from comparing calculations with $x=0$ and $x=1$ are not so obvious. This naturally leads us to the need of looking at the ratio of charged pions in the next subsection.

\subsection{Charged pion ratio as a probe of symmetry energy from central to peripheral reactions}
\begin{figure}[h]
\centerline{\includegraphics[scale=0.52]{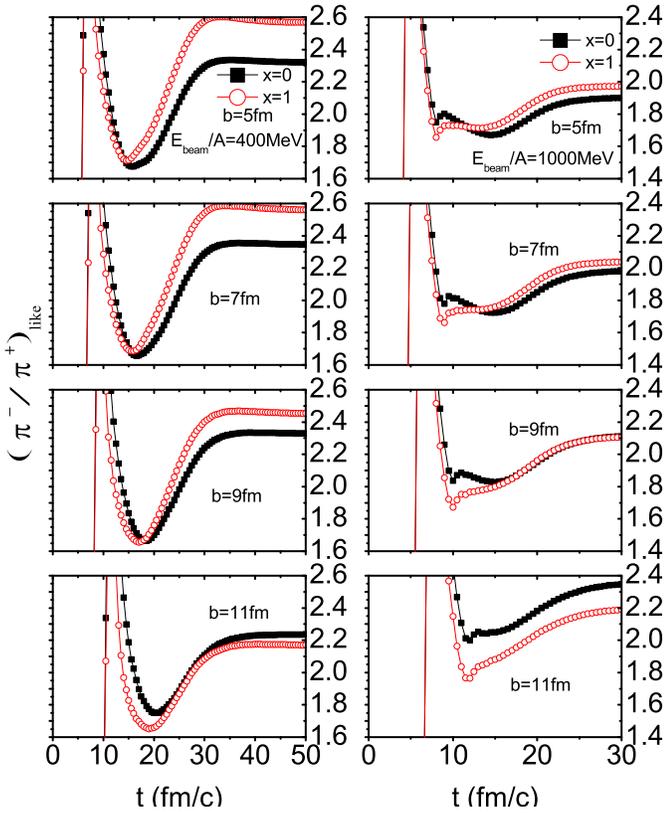}} \caption{(Color
online) Evolution of the $(\pi^{-}/\pi^{+})_{\rm like}$ ratio from
midcentral to peripheral Pb+Pb reactions at beam energies of 400 (left)
and 1000 MeV/nucleon (right). The $x$ parameter is 1 (red) and 0
(black), and the size of neutron skin is 0.1 fm in all cases.} \label{Beam}
\end{figure}
Within the IBUU model for heavy-ion collisions up to about 1.5 GeV/nucleon, most pions are
produced through the decay of $\Delta (1232)$ resonances. As
discussed in detail in Ref. \cite{Bao02}, the
$(\pi^{-}/\pi^{+})_{\rm like}$ ratio defined as
\begin{equation}\label{ratio}
(\pi^{-}/\pi^{+})_{\rm like}\equiv
\frac{\pi^{-}+\Delta^{-}+\frac{1}{3}\Delta^{0}}
{\pi^{+}+\Delta^{++}+\frac{1}{3}\Delta^{+}},
\end{equation}
is a good measure of the symmetry energy during heavy-ion
collisions. At the final stage, all the $\Delta$ resonances will
eventually decay and the $(\pi^{-}/\pi^{+})_{\rm like}$ ratio
naturally becomes the ratio of free pions, i.e., $\pi^{-}/\pi^{+}$.
To see how the \rpi ratio in heavy-ion collisions depends on the
impact parameter, the beam energy, and the \esym for a given neutron
skin thickness of 0.1 fm, we show in Fig. \ref {Beam} the
evolutions of the $(\pi^{-}/\pi^{+})_{\rm like}$ ratio from
midcentral to peripheral $^{208}$Pb+$^{208}$Pb collisions at a beam
energy of 400 (left) and 1000 MeV/nucleon (right), respectively. For
pion production, it is known that there is no obvious impact
parameter dependence from head-on to midcentral heavy-ion reactions.
Consistent with previous observations from various transport model
calculations, the \rpi ratio is larger and more sensitive to the
\esym near the pion production threshold. One interesting new
feature seen is that the dependence of the \rpi ratio on the \esym
has a transition from midcentral to peripheral collisions. Namely,
in central to midcentral collisions, a softer \esym with $x=1$ leads
to a larger \rpi ratio than the stiffer one with $x=0$, while it is
the opposite in peripheral reactions. This is because the average
density of the participant region is significantly above $\rho_0$ in
central to midcentral collisions, while it becomes lower than
$\rho_0$ in peripheral collisions. In the latter case, a softer
\esym gives a larger value of the symmetry energy and a less
neutron-rich participant matter, resulting in a smaller \rpi ratio.
This feature indicates that the sorting of events according to
some effective impact parameter selection criteria is very important for
extracting reliable information about the \esym from the \rpi ratio in
heavy ion collisions.

\subsection{Interplay of neutron skin and symmetry energy on the charged pion ratio}
\begin{figure}[h]
\centerline{\includegraphics[scale=0.37]{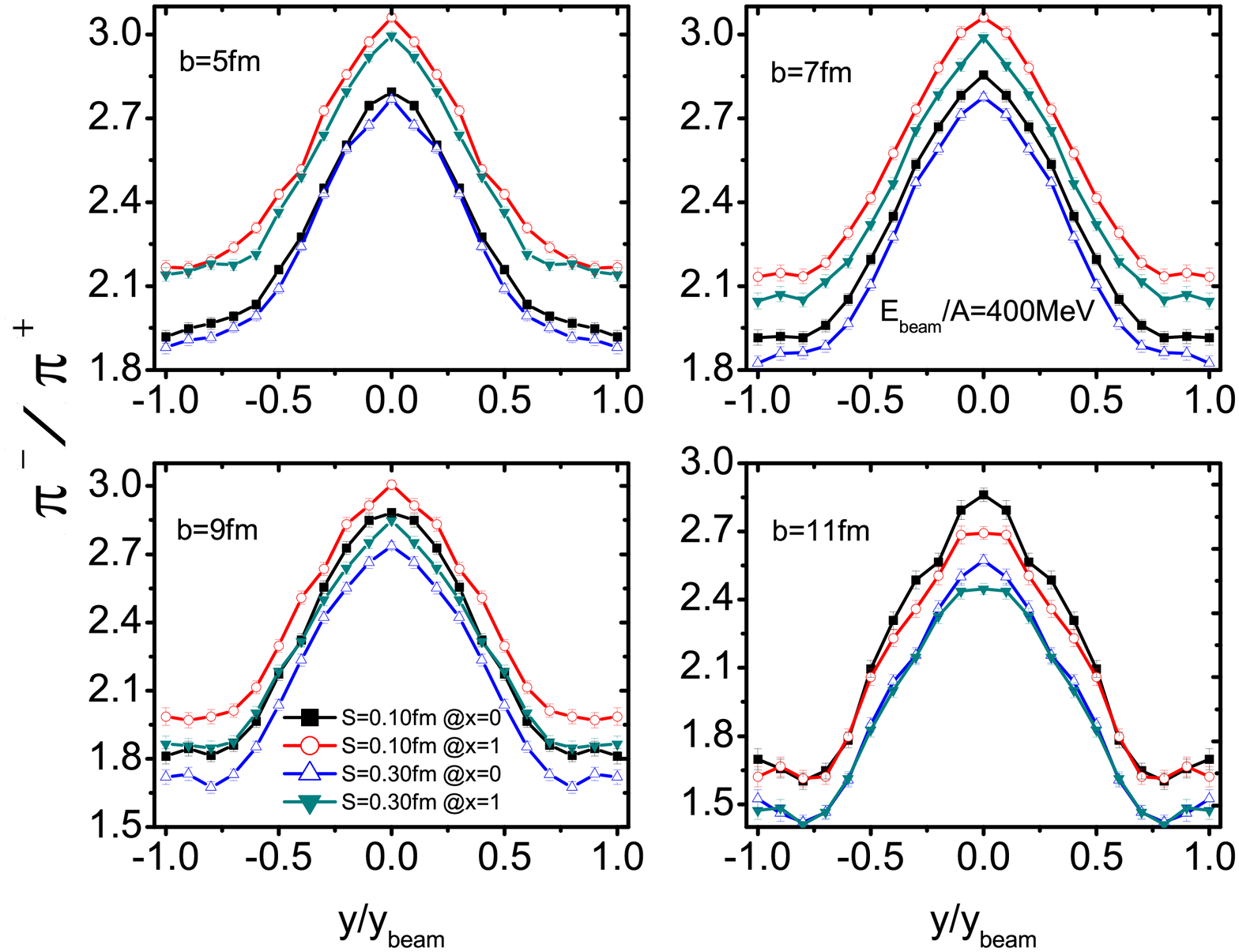}} \caption{(Color
online) Rapidity distributions of the $\pi^{-}/\pi^{+}$ ratio from
midcentral to peripheral Pb+Pb collisions at 400 MeV/nucleon with
two sizes of neutron skin of 0.1 and 0.3 fm.} \label{Impact-1}
\end{figure}

\begin{figure}[h]
\centerline{\includegraphics[scale=0.37]{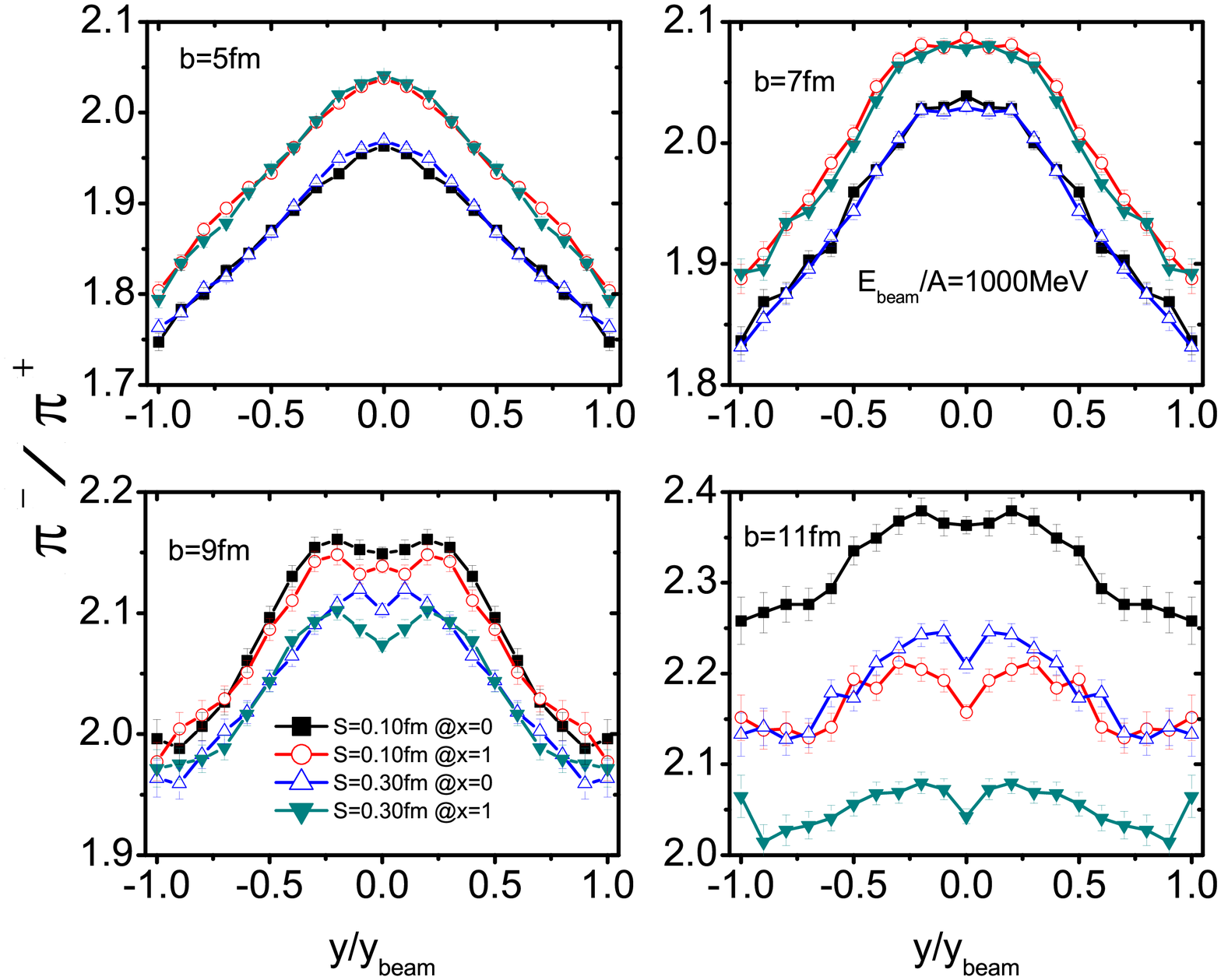}} \caption{(Color
online) Same as Fig.~\ref{Impact-1} but for the beam energy of 1000
MeV/nucleon.} \label{Impact-2}
\end{figure}
\begin{figure}[h]
\centerline{\includegraphics[scale=0.35]{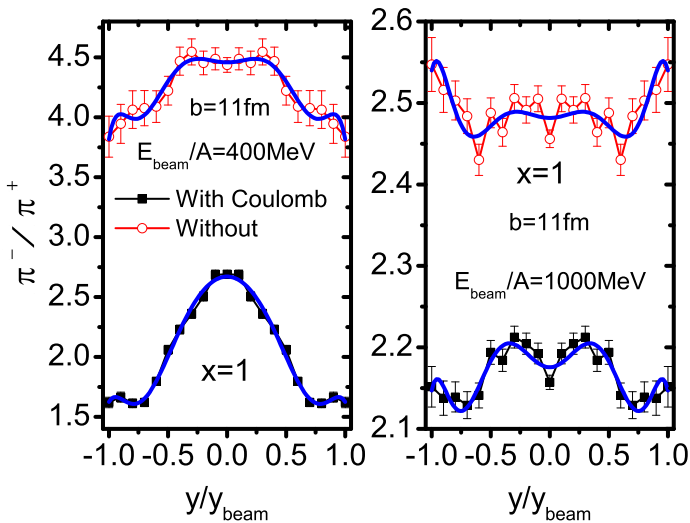}} \caption{(Color
online) Rapidity distributions of the $\pi^{-}/\pi^{+}$ ratio
with the soft symmetry energy (x=1) in Pb+Pb collisions with an impact
parameter of 11 fm and the beam energies of 400 MeV/nucleon and
1000MeV/nucleon with and without the Coulomb fields, respectively. The blue lines are used to guide the eye.} \label{Coulomb}
\end{figure}
\begin{figure}[h]
\centerline{\includegraphics[scale=0.35]{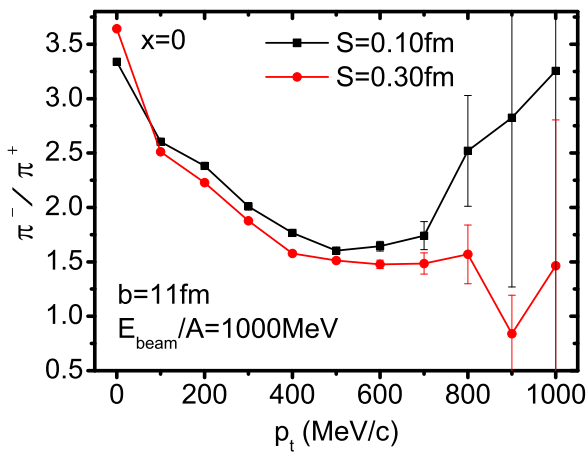}} \caption{(Color
online) Transverse momentum dependence of the $\pi^{-}/\pi^{+}$ ratio
with the stiff symmetry energy (x=0) in Pb+Pb collisions with an impact
parameter of 11 fm and a beam energy of 1000 MeV/nucleon with the
two sizes of neutron skin of 0.1 and 0.3 fm, respectively.} \label{momentum}
\end{figure}
Since the symmetry energy effect depends sensitively on the isospin
asymmetry of the participant region, we would expect that the transitional features shown in Fig.\ \ref{Beam} depend on the size of the
neutron skin and the beam energy. We thus examine the rapidity
dependence of the \rpi ratio using different sizes of
neutron skin in $^{208}$Pb in Fig. \ref{Impact-1} and Fig.
\ref{Impact-2} with beam energy of 400 MeV/nucleon and 1000
MeV/nucleon, respectively. First of all, it is interesting to see again that in central reactions at both beam energies, the charged pion ratio is not much affected by the size of the neutron skin but sensitive to the density dependence of the nuclear symmetry energy. However, the charged pion ratio in peripheral reactions are significantly affected by the size of the neutron skin. Moreover, it is seen that increasing the size of the neutron skin reduces appreciably the \rpi ratio in peripheral
reactions especially at a lower beam energy. Within the first-chance
nucleon-nucleon collision model without considering the isospin
dependence of Pauli blocking as well as the subsequent pion
reabsorption and reproduction that are modeled in the transport
model, one natively expects the \rpi ratio to increase with the increasing size of neutron skin. Indeed, this is observed in the very earlier stage of the reaction.
As shown in Fig. \ref{pmul1}, with a thinner neutron skin the final multiplicities of both $\pi^-$ and $\pi^+$ decrease in peripheral reactions because of the smaller overlap between the two colliding nuclei. However, the multiplicity of $\pi^+$ decreases faster than $\pi^-$, leading to a
larger \rpi value for the reaction with a neutron skin thickness of 0.1 fm. Compared to the 400 MeV/nucleon case, the \rpi ratio at 1000 MeV/nucleon is flatter in rapidity $y$ due to multiple pion production channels possible only at higher beam energies. Moreover, in peripheral reactions at 1000 MeV/nucleon the \rpi curve has a double bump structure. This is understandable since the nuclear stopping power is lower at 1000 than 400 MeV/nucleon. Thus, the Coulomb field of the spectators in peripheral collisions at 1000 MeV/nucleon leads to the well-known Coulomb peaks in the rapidity distribution of the \rpi ratio as shown in Fig. \ref{Coulomb} by comparing simulations with and without including the Coulomb fields.
\begin{table}[h]
\caption{{\protect\small The measure $F(L_{2})$ ($10^{-2}$) of the symmetry
energy effect on the midrapidity $\pi^{-}/\pi^{+}$ ratio in Pb+Pb
reactions.}} \label{tab1}
\begin{tabular}{ccccccc}
\hline\hline
&  $S=0.10$ fm \quad & $S=0.30$ fm \\
\hline
$E_{beam}$(MeV)  &   400 (1000) &400 (1000)\\
$$ $b=5$ fm &  $13.7 ~(4.0) $  & $12.2 ~(3.8) $\\
$$ $b=7$ fm  &  $11.4 ~(3.1) $  & $10.4 ~(2.9) $\\
\hline\hline
\end{tabular}%
\end{table}
\begin{table}[h]
\caption{{\protect\small The measure $F(S)$ ($10^{-2}$) of the neutron
skin effect on the midrapidity $\pi^{-}/\pi^{+}$ ratio in Pb+Pb
reactions.}} \label{tab2}
\begin{tabular}{ccccccc}
\hline\hline
  &  $x=0$  \quad & $x=1$ \\
\hline
$E_{beam}$ (MeV)  &   400 (1000) &400 (1000)\\

$$ $b=5$ fm &  $4.8 ~(0.4) $  & $8.9 ~(0.2) $\\
$$ $b=7$ fm  &  $11.0 ~(0.4) $  & $13.9 ~(0.9) $\\
$$ $b=9$ fm &  $22.1 ~(7.8)$  & $25.2 ~(7.2)$\\
$$ $b=11$ fm  &  $41.8 ~(20.6)$  & $33.4 ~(18.5)$\\
\hline\hline
\end{tabular}%
\end{table}

As shown in Fig. \ref{Impact-1} and Fig. \ref{Impact-2}, while the
absolute effect of the neutron skin is small especially in central reactions, it is comparable or
even larger than the effect due to the \esym in peripheral
collisions, especially at 400 MeV/nucleon. Considering that one of the main
interesting aspects of \rpi ratio is its sensitivity to the nuclear symmetry energy at suprasaturation densities
in central heavy-ion collisions while in peripheral reactions it is
sensitive to the size of the neutron skin, we compare more quantitatively
their relative effects on the \rpi ratio by using
\begin{equation}\label{F(L_{2})}
F(L_{2})=\frac{\Delta(\pi^{-}/\pi^{+})}{{\Delta L_{2}}/{L_{2}}}
\end{equation}
and
\begin{equation}\label{F(S)}
F(S)=\frac{\Delta(\pi^{-}/\pi^{+})}{{\Delta S}/S},
\end{equation}
where $\Delta (\pi^{-}/\pi^{+})$ is the resulting change in the
midrapidity ($|y/y_{beam}|\leqslant 0.5$) \rpi ratio due to the variation of $L_2$
or the size of the neutron skin thickness $S$ in the initial state. Obviously, for the considered reactions only with
impact parameters less than about 7 fm, the central density can reach twice the normal density for a
short time. We thus only show the $F(L_2)$ for central reactions in Table \ref{tab1}. It is seen that the
values of $F(L_2)$ are about the same with both $S=0.1$ and 0.3 fm. They are indeed significantly larger than the corresponding values of $F(S)$ shown in Table \ref{tab2}, indicating that
indeed the \rpi ratio in these central reactions is an observable sensitive
to the high-density symmetry energy with little influence from the uncertainties of
the size of neutron skin. On the other hand, in more peripheral
collisions the \rpi ratio is sensitive to the variation of neutron
skin thickness with small uncertainties due to our poor knowledge
about the suprasaturation \esym, especially at lower beam energies.
Moreover, we expect that pions with higher transverse momenta emitted from the
skin-skin interaction region have more chances to escape. Shown in Fig. \ref{momentum} is the
transverse momentum dependence of the \rpi ratio in peripheral
Pb+Pb collision with an impact parameter of b=11 fm and a beam energy of 1000 MeV/nucleon.
Indeed, the \rpi ratio with higher transverse momenta is more sensitive to the neutron skin thickness in peripheral reactions.

\section{Summary}
In summary, the influence of the uncertainty of the neutron skin
thickness on the $\pi^{-}/\pi^{+}$ ratio in $^{208}$Pb+$^{208}$Pb
collisions at beam energies of 400 and 1000 MeV/nucleon was examined
within the IBUU11 transport model. While the neutron skin effect on
the \rpi ratio is negligible in central reactions at both energies,
it increases gradually with increasing impact parameter and becomes
comparable with or even larger than the symmetry energy effect in
peripheral collisions especially at 400 MeV/nucleon. Moreover, it is
found that while the \rpi ratio is higher with a softer \esym in
central collisions, above certain impact parameters depending on the
size of neutron skin a stiffer \esym can lead to a larger \rpi
ratio. Therefore, a clear impact parameter
selection is important to extract reliable information about the
\esym at suprasaturation densities (size of neutron skin)
from the $\pi^-/\pi^+$ ratio in central (peripheral) heavy-ion collisions.\\
\\
\noindent{\textbf{Acknowledgements}} \\
We would like to thank Farrooh Fattoyev,
Wei-Zhou Jiang, William G. Newton, and Li Ou for helpful discussions
and their strong support in various ways. We also thank the help
provided by the supporting staff of the High-Performance
Computational Science Research Cluster at Texas A$\&$M
University-Commerce where all the calculations were done. This work
was supported in part by the US National Science Foundation grants
PHY-1068022, the National Aeronautics and Space Administration under
grant N- NX11AC41G issued through the Science Mission Directorate,
the CUSTIPEN (China-U.S. Theory Institute for Physics with Exotic
Nuclei) under DOE grant number DE-FG02-13ER42025, the NNSF of China
(11320101004, 11135011, 11275125, 11035009, and 11220101005), the Shanghai
Rising-Star Program (11QH1401100), Shanghai ``Shu Guang" Project, the
Eastern Scholar Program, the STC of Shanghai Municipality
(11DZ2260700), and the ``100-talent plan" of Shanghai Institute of
Applied Physics under grant Y290061011 from the Chinese Academy of
Sciences.

\end{document}